\PassOptionsToPackage{unicode}{hyperref}
\PassOptionsToPackage{hyphens}{url}
\documentclass[
]{article}
\usepackage{amsmath,amssymb}
\usepackage{iftex}
\ifPDFTeX
  \usepackage[T1]{fontenc}
  \usepackage[utf8]{inputenc}
  \usepackage{textcomp} 
\else 
  \usepackage{unicode-math} 
  \defaultfontfeatures{Scale=MatchLowercase}
  \defaultfontfeatures[\rmfamily]{Ligatures=TeX,Scale=1}
\fi
\usepackage{lmodern}
\ifPDFTeX\else
\fi
\IfFileExists{upquote.sty}{\usepackage{upquote}}{}
\IfFileExists{microtype.sty}{
  \usepackage[]{microtype}
  \UseMicrotypeSet[protrusion]{basicmath} 
}{}
\makeatletter
\@ifundefined{KOMAClassName}{
  \IfFileExists{parskip.sty}{%
    \usepackage{parskip}
  }{
    \setlength{\parindent}{0pt}
    \setlength{\parskip}{6pt plus 2pt minus 1pt}}
}{
  \KOMAoptions{parskip=half}}
\makeatother
\usepackage{xcolor}
\usepackage[margin=1in]{geometry}
\providecommand{\tightlist}{%
  \setlength{\itemsep}{0pt}\setlength{\parskip}{0pt}}
\newlength{\cslhangindent}
\newlength{\csllabelwidth}
\newlength{\cslentryspacingunit} 
\newenvironment{CSLReferences}[2] 
 {
  \setlength{\parindent}{0pt}
  \ifodd #1
  \let\oldpar\par
  \def\par{\hangindent=\cslhangindent\oldpar}
  \fi
  \setlength{\parskip}{#2\cslentryspacingunit}
 }%
 {}
\usepackage{calc}

\newcommand{\CSLLeftMargin}[1]{\parbox[t]{\csllabelwidth}{#1}}
\newcommand{\CSLRightInline}[1]{\parbox[t]{\linewidth - \csllabelwidth}{#1}\break}

\usepackage{graphicx}
\ifLuaTeX
  \usepackage{selnolig}  
\fi
\IfFileExists{bookmark.sty}{\usepackage{bookmark}}{\usepackage{hyperref}}
\IfFileExists{xurl.sty}{\usepackage{xurl}}{} 
\hypersetup{
  pdftitle={Public-key encryption from a trapdoor one-way embedding of SL\_2(\textbackslash mathbb\{N\})},
  hidelinks,
  pdfcreator={LaTeX via pandoc}}

\title{Public-key encryption from a trapdoor one-way embedding of
\(SL_2(\mathbb{N})\)}
\author{Robert Hines\\
\texttt{alphanumericnonsense@gmail.com}}
\date{\today}

\begin{document}
\maketitle
\begin{abstract}
We obfuscate words of a given length in a free monoid on two generators
with a simple factorization algorithm (namely \(SL_2(\mathbb{N})\)) to
create a public-key encryption scheme. We provide a reference
implementation in Python and suggested parameters. The security analysis
is between weak and non-existent, left to future work.
\end{abstract}

\hypertarget{introduction}{%
\section{Introduction}\label{introduction}}

The idea for the encryption scheme is straightforward. First map
bitstrings of length \(\lambda\) to products of length \(\lambda\) in a
free monoid on two generators. The particular monoid instance should
have a simple factorization algorithm. Next, homomorphically obfuscate
the monoid (trapdoor one-way embedding) and provide the obfuscated
generators as a public key. We provide a detailed algorithmic
instantiation of this idea below with the monoid \(SL_2(\mathbb{N}\))
and a Python implementation of the resulting scheme for experimentation.

While there is much to be done concerning cryptanalysis, we give a list
of important questions affecting security and practicality of the
scheme.

\hypertarget{prior-art}{%
\section{Prior art}\label{prior-art}}

An existing ``non-commutative knapsack'' approach overlapping
conceptually with our proposal is
\protect\hyperlink{ref-matrix}{{[}1{]}}. Our use of the algebraic
structure of \(SL_2(\mathbb{N})\) (as opposed to trial-and-error and
heuristics) and smaller public keys (two generators instead of a
knapsack setup) is arguably more elegant if nothing else.

\hypertarget{the-medium-sl_2mathbbn}{%
\section{\texorpdfstring{The medium:
\(SL_2(\mathbb{N})\)}{The medium: SL\_2(\textbackslash mathbb\{N\})}}\label{the-medium-sl_2mathbbn}}

Let \(L=\left(\begin{array}{cc}1&0\\1&1\\\end{array}\right)\) and
\(R=\left(\begin{array}{cc}1&1\\0&1\\\end{array}\right)\). These two
matrices generate \(SL_2(\mathbb{N})\) as free monoid on two generators,
i.e.~every \(2\times2\) matrix with non-negative integer entries and
determinant one can be written uniquely as a finite word in the alphabet
\(\{L,R\}\). We note for later use that a product/word of length \(k\)
in \(\{L,R\}\) has non-negative integer entries bounded by \(m=2^k\).
This bound could be made smaller, e.g.~the \(k\)th Fibonacci number
\(F_k\) is the maximum matrix entry of words of length \(k\), but the
bound \(m=2^k\) will suffice for our purposes.

Moreover, given an element
\(M=\left(\begin{array}{cc}a&b\\c&d\\\end{array}\right)\in SL_2(\mathbb{N})\),
we can recover the factorization using some variant of the Euclidean
algorithm or continued fractions. For instance:

\begin{itemize}
\tightlist
\item
  If \(a/c\leq 1\) then \(M=LM'\), and if \(a/c>1\) (including
  \(1/0=\infty\)), then \(M=RM'\).
\item
  Left multiply by \(L^{-1}\) or \(R^{-1}\) and continue until \(M'=1\).
\end{itemize}

\hypertarget{the-trapdoor-modular-reduction-conjugation-increasing-dimension-and-random-generators.}{%
\section{The trapdoor: modular reduction, conjugation, increasing
dimension, and random
generators.}\label{the-trapdoor-modular-reduction-conjugation-increasing-dimension-and-random-generators.}}

We start with a naive attempt at a trapdoor and describe refinements to
obtain something useful (or so we hope).

\hypertarget{first-attempt-conjugate}{%
\subsection{First attempt (conjugate)}\label{first-attempt-conjugate}}

Denote by \(SL_2^{(\lambda)}(\mathbb{N})\) the set of all words of
length \(\lambda\) in the alphabet \(\{L,R\}\). As a first attempt to
obfuscate \(M\in SL_2^{\lambda}(\mathbb{N})\), we work modulo \(m\),
where \(m\) is any integer greater than
\(\max\{\|A\|_{\infty} : A\in SL_2^{(\lambda)}(\mathbb{N})\}\), and
conjugate by a uniformly random element
\(S\in GL_2(\mathbb{Z}/m\mathbb{Z})\), \(M\mapsto S^{-1}MS\).

However, \(\{S^{-1}LS, S^{-1}RS\}\) leaks too much information, namely
the algebraic relations \[
S^{-1}LS=\frac{1}{ad-bc}\left(\begin{array}{cc}-ab+ad-bc&-b^2\\a^2&ab+ad-bc\\\end{array}\right),
\] and \[
S^{-1}RS=\frac{1}{ad-bc}\left(\begin{array}{cc}ad-bc+cd&d^2\\-c^2&ad-bc-cd\\\end{array}\right),
\] where \(S=\left(\begin{array}{cc}a&b\\c&d\\\end{array}\right)\).
These are enough to recover the secret \(S\) by finding square roots
modulo \(m\).

\hypertarget{strike-two-increase-the-dimension}{%
\subsection{Strike two (increase the
dimension)}\label{strike-two-increase-the-dimension}}

We can first increase the dimension, setting a parameter \(n\geq 2\),
embedding \[
L\mapsto \widetilde{L}=\left(\begin{array}{cc}I_n&0\\I_n&I_n\\\end{array}\right), \quad R\mapsto \widetilde{R}=\left(\begin{array}{cc}I_n&I_n\\0&I_n\\\end{array}\right),
\] and conjugating by a secret \(S\in GL_{2n}(\mathbb{Z}/m\mathbb{Z})\).
By taking \(n>1\), we get equations similar to the first attempt, \[
S^{-1}\widetilde{L}S=\left(\begin{array}{cc}EA+FA+FC&EB+FB+FD\\GA+HA+HC&GB+HB+HD\\\end{array}\right),
\] and \[
S^{-1}\widetilde{R}S=\left(\begin{array}{cc}EA+EC+FC&EB+ED+FD\\GA+GC+HC&GB+GD+HD\\\end{array}\right),
\]

where \(S=\left(\begin{array}{cc}A&B\\C&D\\\end{array}\right)\),
\(S^{-1}=\left(\begin{array}{cc}E&F\\G&H\\\end{array}\right)\). One
would hope that these are more difficult to solve, but perhaps
cryptographically secure values of \(n\) would make public key and
ciphertext sizes impractical.

\hypertarget{third-times-a-charm-use-random-generators}{%
\subsection{Third time's a charm (use random
generators)}\label{third-times-a-charm-use-random-generators}}

Instead of using the generators \(\{L,R\}\), we could randomly choose
\(\{G_0,G_1\}\), each a word of length \(l\) in \(\{L,R\}\),
i.e.~\(\{G_0,G_1\}\subseteq SL_2^{(l)}(\mathbb{N})\). For large enough
\(l\), the resulting public key and plain/cipher pairs should be secure.
The cost to pay for this is an increase in the bound \(m\), namely we
must work with \(SL_2^{(l\lambda)}(\mathbb{N})\) and take something like
\(m=2^{l\lambda}\). For practicality, \(l\) should be on the small side.

\hypertarget{four-all-of-the-above}{%
\subsection{Four (all of the above)?}\label{four-all-of-the-above}}

Finally, we could combine the second and third attempts by using random
generators and larger matrices. Whether or not this is prudent depends
on a more careful analysis of the weaknesses of each.

\hypertarget{public-key-encryption-scheme}{%
\section{Public key encryption
scheme}\label{public-key-encryption-scheme}}

In the following, we choose parameters \(\lambda\), \(l\), and \(m\),
and let \(G_0\), \(G_1\) be obtained from randomly selected words of
length \(l\) in \(\{L,R\}\).

\hypertarget{key-generation}{%
\subsection{Key generation}\label{key-generation}}

The secret key consists of \(S\), a uniformly random element of
\(GL_{2n}(\mathbb{Z}/m\mathbb{Z})\), and the random matrices
\(\{G_0,G_1\}\). The public key is the pair
\(P_0=S^{-1}\widetilde{G_0}S\), \(P_1=S^{-1}\widetilde{G_1}S\), with all
arithmetic operations over \(\mathbb{Z}/m\mathbb{Z}\).

\hypertarget{encryption}{%
\subsection{Encryption}\label{encryption}}

Let \(\mu\) be a \(\lambda\)-bit message. To encrypt
\(\mu=\mu_0\mu_1\cdots\mu_{\lambda-1}\), form the product
\(C=\prod_{i=0}^{\lambda-1}P_{\mu_i}\in SL_{2n}(\mathbb{Z}/m\mathbb{Z})\),
with all arithmetic modulo \(m\).

\hypertarget{decryption}{%
\subsection{Decryption}\label{decryption}}

To decrypt the ciphertext \(C\), first conjugate by \(S^{-1}\) to get
\(SCS^{-1}=\prod_{i=0}^{\lambda-1}\widetilde{G_{\mu_i}}\) (arithmetic
modulo \(m\)). This is now an unobfuscated element of
\(SL_2(\mathbb{N})\) (after reducing a block ``constant''
\(2n\times 2n\) matrix to a \(2\times 2\) matrix). This can be factored
easily as described in an earlier section to recover \(\mu\).

\hypertarget{security-and-parameter-selection}{%
\section{Security and parameter
selection}\label{security-and-parameter-selection}}

Let \(F_2^{(\lambda)}\) be the set of words of length \(\lambda\) in the
free monoid on two generators. Is the embedding defined by the public
key,
\(\epsilon_{P_0, P_1}:F^{(\lambda)}_2\hookrightarrow SL_{2n}(\mathbb{Z}/m\mathbb{Z})\),
one-way? While this is the paramount security question, we leave it
aside for the moment and consider potential practicality first. We have
\[
|\mathsf{sk}|=4n^2l\lambda+2l, \quad |\mathsf{pk}|=8n^2l\lambda, \quad |\mathsf{ct}|=4n^2l\lambda.
\] Another concern is the large \(l\lambda\)-bit integer arithmetic. As
a starting point, we could take \(l=\lambda=256\) and \(n=1\), naive
256-bit security against finding the generators or plaintext recovery
with 256-bit plaintexts. While key and ciphertext sizes are
``reasonable,'' the \(2^{16}\)-bit multiplication is not. Sliding the
parameters to \(n=16\), \(l=1\), \(\lambda=256\) keeps the same key and
ciphertext sizes and reasonable 256-bit arithmetic. A parameter choice
in between the two extremes is \(l=16\), \(n=4\), \(\lambda=256\) (again
keeping key and ciphertext sizes fixed), with 4096-bit multiplication
(in line with modern RSA). At this point considering smaller bounds on
\(m\) starts looking good, with smaller exponential growth in \(m\) as a
function of \(\lambda\).

\hypertarget{attacking-conjugation}{%
\subsection{Attacking conjugation}\label{attacking-conjugation}}

Suppose \(n=1\) for simplicity. The public key \((P_0,P_1)\) and
ciphertext \(C\) don't hide the conjugacy classes of \((G_0, G_1)\) or
\(\mu\) encoded as a product of \(\{L,R\}\) matrices, i.e.~the trace is
preserved or almost preserved modulo \(m\). The \(2\times 2\) matrix
\(M=\left(\begin{array}{cc}a&b\\c&d\\\end{array}\right)\) we're looking
for (say one of \(G_0\), \(G_1\), or \(\mu\)) is a solution to \[
\left\{
\begin{array}{c}
ad-bc = 1,\\
a+d=t \quad (t \text{ is the trace)},\\
0\leq a,b,c,d<m=2^k, \text{ e.g. } 2^l \text{ or } 2^{l\lambda},\\
\mathsf{len}_{\{L,R\}}(M)=k,\text{ e.g. } l \text{ or } l\lambda,\\
M':=S^{-1}MS \bmod m=2^k\text{ is random and known},\\
\end{array}
\right.
\] where \(t\) is the known trace. To the best of our knowledge, these
are essentially the constraints, although some number theory may help
(\(M\) is probably hyperbolic, stabilizes its fixed point form
\(\pm F(x,y)=cx^2+(d-a)xy-by^2\) of discriminant \(D=t^2-4\), the class
number of \(\mathbb{Q}(\sqrt{D})\) is probably 1, Pell's equation is
easy to solve via continued fractions, etc.).

In other words the first problem is:

\begin{quote}
Given a random element \(G=S^{-1}MS\) (\(S\) uniformly random on
\(GL_2(\mathbb{Z}/m\mathbb{Z})\), \(m=2^k\)) of the conjugacy class of
an element \(M\in SL_2^{(k)}(\mathbb{N})\) (i.e.~length \(k\) in
\(\{L,R\}\)), what is the complexity of determining \(M\) as a function
of \(k\)?
\end{quote}

Solving the problem above with \(k=l\lambda\) suffices for plaintext
recovery. It is also a first step towards secret recovery, i.e.~finding
\(G_0\) and/or \(G_1\) from \(P_0=S^{-1}G_0S\), \(P_1=S^{-1}G_1S\)
before solving for the secret \(S\).

Another question regarding the conjugacy class search is:

\begin{quote}
What is the distribution of \(\mathsf{tr}(M)\) for
\(M\in SL_2^{(k)}(\mathbb{N})\)?
\end{quote}

For instance, if this isn't uniform enough on its support (i.e.~traces
are biased on length \(k\) words), that could weaken the cryptosystem if
exploitable. There are obvious non-uniformities, e.g.~\(L^k\) and
\(R^k\) are the only parabolics (trace 2).

\begin{figure}[h]
\begin{center}
\includegraphics[width=0.3\textwidth]{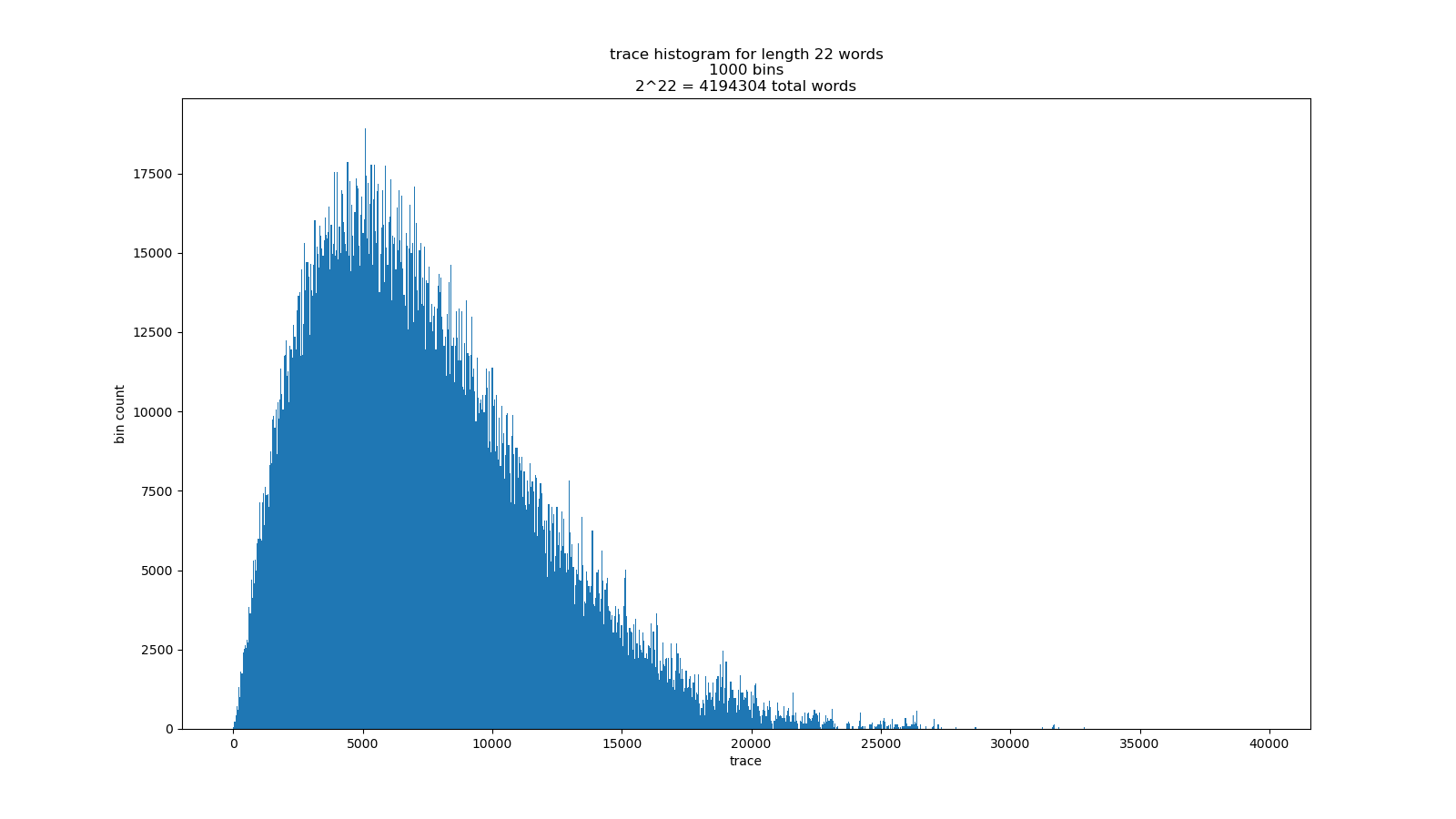}
\includegraphics[width=0.3\textwidth]{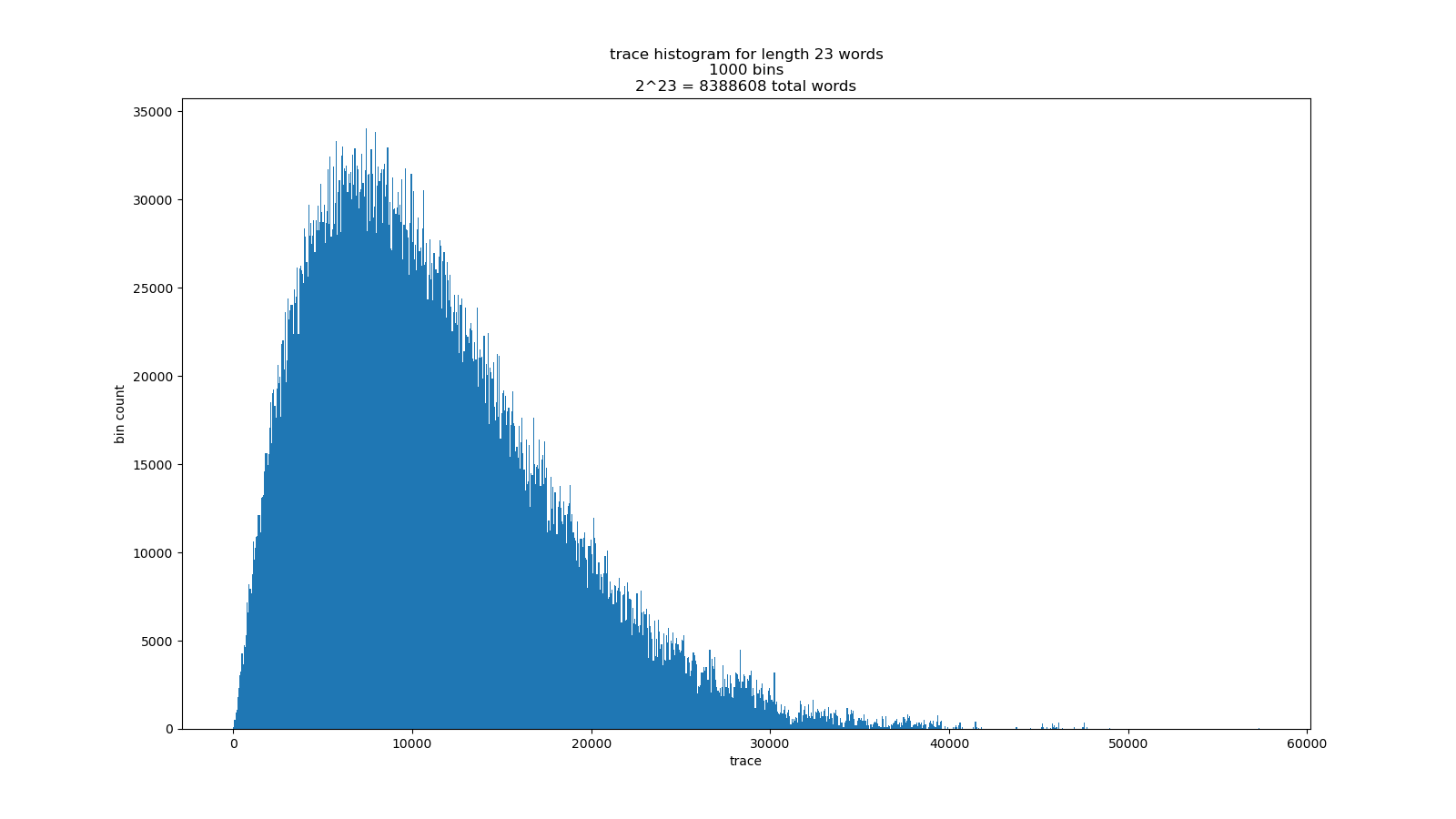}
\includegraphics[width=0.3\textwidth]{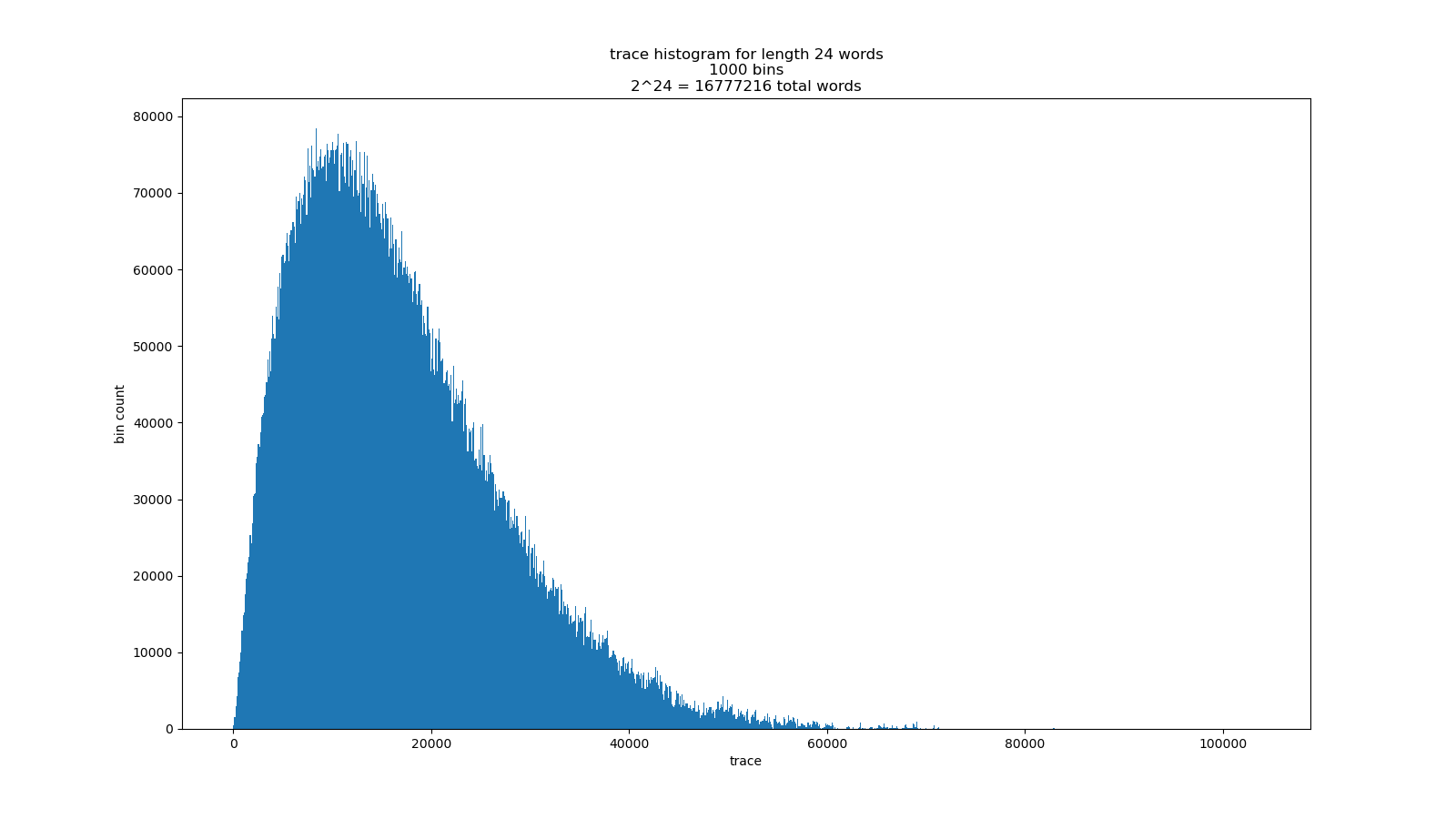}
\caption{Trace distributions for words of length 22, 23, and 24. Histograms with 1000 bins of equal size.}
\label{fig:tracedist}
\end{center}
\end{figure}

The trace distribution may be explicitly computable; there's a sort of
``non-commutative'' recurrence in the Cayley tree of the monoid with
respect to the generators \(\{L, R\}\), namely \[
\mathsf{tr}(ML)=\mathsf{tr}(M)+b, \quad \mathsf{tr}(MR)=\mathsf{tr}(M)+c
\] where \(M=\left(\begin{array}{cc}a&b\\c&d\\\end{array}\right)\).
Experimentally, \autoref{fig:tracedist} shows distributions of traces
for words of length \(k=22\), \(23\), and \(24\).

\hypertarget{attacking-the-secret-key-s}{%
\subsection{\texorpdfstring{Attacking the secret key
\(S\)}{Attacking the secret key S}}\label{attacking-the-secret-key-s}}

Suppose we have known or chosen plain or ciphertext, or better yet the
generators \(G_0\), \(G_1\) (which is the case with \(l=1\)). When
\(n=1\), we've already seen the resulting system of equations is easy to
solve, taking square roots modulo \(m\). Does increasing \(n\) make the
problem more difficult?

\begin{quote}
What is the computational complexity of finding \(S\) from known \(M\)
and \(S^{-1}\widetilde{M}S\) as a function of \(n\)?
\end{quote}

Using the same notation as before (\(n\times n\) blocks for \(S\),
\(S^{-1}\)), \[
S=\left(\begin{array}{cc}A&B\\C&D\\\end{array}\right), \quad S^{-1}=\left(\begin{array}{cc}E&F\\G&H\\\end{array}\right), \quad M=\left(\begin{array}{cc}a&b\\c&d\\\end{array}\right),
\] we have \[
S^{-1}\widetilde{M}S=\left(\begin{array}{cc}aEA+cFA+bEC+dFC&aEB+cFB+bED+dFD\\aGA+cHA+bGC+dHC&aGB+cHB+bGD+dHD\\\end{array}\right).
\] With minor assumptions, we can write expressions for \(S^{-1}\) in
terms of \(S\) (cf. \protect\hyperlink{ref-blockinv}{{[}2{]}}), e.g.~if
\(A\) is invertible then \((D-CA^{-1}B)\) is invertible and \[
S^{-1}=\left(\begin{array}{cc}A^{-1}+A^{-1}B(D-CA^{-1}B)^{-1}CA^{-1}&-A^{-1}B(D-CA^{-1}B)^{-1}\\-(D-CA^{-1}B)^{-1}CA^{-1}&(D-CA^{-1}B)^{-1}\\\end{array}\right).
\] For simplicity, say \(a=1\), \(b=c=d=0\), so that \[
S^{-1}\widetilde{[1,0;0,0]}S=
\left(\begin{array}{cc}
I+A^{-1}B(D-CA^{-1}B)^{-1}C
&A^{-1}+A^{-1}B(D-CA^{-1}B)^{-1}CA^{-1}B\\
(D-CA^{-1}B)^{-1}C
&(D-CA^{-1}B)^{-1}CA^{-1}B\\
\end{array}\right).
\] This still seems formidable, with \(\approx 4n^2\) independent
variables and \(4n^2\) equations, with a solution determined up to the
centralizer of \(M\).

\hypertarget{implementation-example-challenge}{%
\section{Implementation, example,
challenge}\label{implementation-example-challenge}}

\hypertarget{implementation}{%
\subsection{Implementation}\label{implementation}}

A Python implementation is provided (\texttt{sl2\_pke.py}\footnote{\url{https://github.com/alphanumericnonsense/sl2_pke}}),
tested at the first and third parameter sets from the previous section.
The implementation is not optimized so takes a while for the suggested
parameters. (It seems large \(n\) is a computational barrier for our
implementation.) The times listed include key generation, encryption,
and decryption on a typical desktop computer (function \texttt{test} in
\texttt{sl2\_pke.py}).

\begin{enumerate}
\def\labelenumi{\arabic{enumi}.}
\tightlist
\item
  \(l=\lambda=256\), \(n=1\): 19.68 seconds (average over 100 trials).
\item
  \(l=1\), \(\lambda=256\), \(n=16\): \(\infty\) (i.e.~more than I
  wanted to wait).
\item
  \(l=16\), \(\lambda=256\), \(n=4\): 14.26 seconds (average over 100
  trials).
\end{enumerate}

In the implementation, the secret key consists of binary \(\{L,R\}\)
representations of \(G_0\), \(G_1\), along with \(S\), \(S^{-1}\) so as
to avoid additional computation during decryption. The public key is
just \((P_0,P_1)\). For all three parameter sets we have \[
|\mathsf{pk}|=8n^2l\lambda=2^{19} \text{ bits}, \quad |\mathsf{ct}|=4n^2l\lambda=2^{18} \text{ bits}.
\]

The bulk of the time is spent in key generation, computing determinants
and inverses (rejection sampling that \(\det(S)\) is a unit and \(S\) is
invertible). The implementation computes determinants recursively using
expansion along the first row, and computes inverses via the adjugate.
One could work with \(m\) prime to increase the proportion of invertible
\(S\) and reduce the number of iterations in the rejection sampling.

\hypertarget{example}{%
\subsection{Example}\label{example}}

For a small example, let's take \(l=8\), \(\lambda=16\), and \(n=2\).

\begin{verbatim}
>>> import sl2_pke
>>> sl2_pke.test(8,16,2,verbose=True) 

parameters: l=8, lmbd=16, n=2 
elapsed time: 0.003463268280029297 seconds 

secret G0bin = [0, 1, 0, 1, 1, 1, 0, 1] 
secret G1bin = [0, 1, 1, 0, 1, 1, 1, 0] 
secret S = 
[[180044238861648907639530005239728443402, 78671938110503256356535945459493689260,
328208603043518655453836457868382331520, 189858830372659235000974137164677995131], 
[155672343735660761206347097294905535103, 106372412637255346718676377918699425235, 
90691651139081347774838965428357027578, 227922651931749744248605342978520132122], 
[3094968359866844574057602930186763998, 266641831191564483984905159250413954559, 
20998567957940753941110035998801658139, 160767525724397655375861666686953163840], 
[178211095150397608370330506792434650267, 180684221329008827603458918903876565245, 
139118072370641183190339253932365801613, 216052035797160283495565699148439592169]] 
secret Sinv = 
[[307192635118851668906543466683872777855, 269289625030035436706852480597147233360, 
159981812487302740355310113004898255387, 53408120534613588930947727648750510691], 
[112181663788938666557979307486052261417, 102697689286225385400371584388074308505, 
322861037399008184822766002568529216261, 174054502788728153025050567413076609419], 
[237808841455499161862946284947239379397, 223110780377640152845753329325645361211, 
146944259742648249323012403503666467604, 294481618604031787995793308484031011747], 
[251449403454450419481736276209944042317, 50503291454136851494286336169449176188, 
306319459087838687149182537269619092194, 230711238969461627876606261311237484890]] 

public P0 = 
[[137413500478426757237729785498897169715, 231239968320834077426658982689680266534, 
69146495480420536278875582615170210772, 33152708537018489215321077094893980946], 
[330162081674891111389361987769479757090, 191423324484097502089583299931615008607, 
128679066393417423397836870943530733124, 275709108234098948675338258762948236678], 
[88563254399421030625099691365843525254, 77201315085951278687430362820053441998, 
95881203497923166664751994007793402653, 278369838638176645352618961861731543340], 
[36846395712098519519183847686723445740, 173455529802750487353281038241693342532, 
244424815999490852163822120880693491112, 255846705381429500934684135425230842005]] 
public P1 = 
[[247715784565787532463649675348437760736, 5908254668445566063898258570842489086,
324033841401195836787414966110816031953, 52100666627039871752387760945836011664], 
[253980257488570419290325329821150515583, 258476128188832362048216706929886240643, 
267003341997740339329266016284380686648, 132860127655589835033045029152588691757], 
[251124012141290027957796299197635681645, 221946805871930782704353653080782150876, 
230415952009162124747908556224420704349, 308271111609761317369148020906687311082], 
[103786741178691326012204819233272678118, 160618027061409062154728853652258911287, 
299106533718846902032063526114038025197, 284239235999033371130348883792559928698]] 

mu = 0xa7b4 
C =
[[84846882633485346302415687463540406713, 109823726553789133715145299719758574752, 
195790665770780782515747872767967575476, 312932876026461235025803616950559157428], 
[45970784775214445249552824804673030496, 138292064403972714254128143616197175537, 
28647030097311286502339440179950584520, 52870673224418264207541283300288313272], 
[218266962851451516080547949287673032976, 16944718053916541257954640898524170932, 
62841792361214906474228062687628559093, 204026872879168090956118039963279146212], 
[29698985555653754674911527304272769264, 55142184405517012638086945268061963020, 
24268175365771347027333132423631575636, 54301627522266814089372838268339482341]] 
mu_recovered = 0xa7b4 

SUCCESS
\end{verbatim}

\hypertarget{challenge}{%
\subsection{Challenge}\label{challenge}}

Data for the first and third parameter sets can easily be obtained from
the given implementation.

\begin{enumerate}
\def\labelenumi{\arabic{enumi}.}
\tightlist
\item
  \(l=\lambda=256\), \(n=1\),
\item
  \(l=1\), \(\lambda=256\), \(n=16\) (bad implementation or
  computationally infeasible),
\item
  \(l=16\), \(\lambda=256\), \(n=4\).
\end{enumerate}

We welcome any attempt to break these (or other) parameter sets, or any
further analysis of the scheme.

\hypertarget{a-final-improvement-not-implemented}{%
\section{A final improvement (not
implemented)}\label{a-final-improvement-not-implemented}}

In light of some of the observed limitations, (large integer arithmetic,
non-uniform trace distribution), and with an eye towards a key
encapsulation mechanism, one potential improvement is to produce
\(G_0\), \(G_1\), and \(\mu\) with traces uniformly in the ``bulk'' of
the trace distribution and use an appropriately smaller value of \(m\).
The price to pay is an increase in the size of \(l\) and \(\lambda\).

For instance, if some large fraction \(f\) of the messages are in the
intersection \[
\{M:t_0\leq\mathsf{tr}(M)\leq t_1\}\cap\{M : \|M\|_{\infty}\leq m_{red}\},
\] say with traces in an interval around the trace \(t_{max}\) at which
the peak density occurs and \(m_{red}\) around half that value, this
should result in a small loss in the message space (i.e.~a small
increase in \(l\), \(\lambda\)) with a much smaller reduction constant
\(m_{red}\ll m\approx 2^n\). In other words, the benefits could outweigh
the cost to give a more practical scheme.

\begin{figure}[h]
\begin{center}
\includegraphics[width=0.3\textwidth]{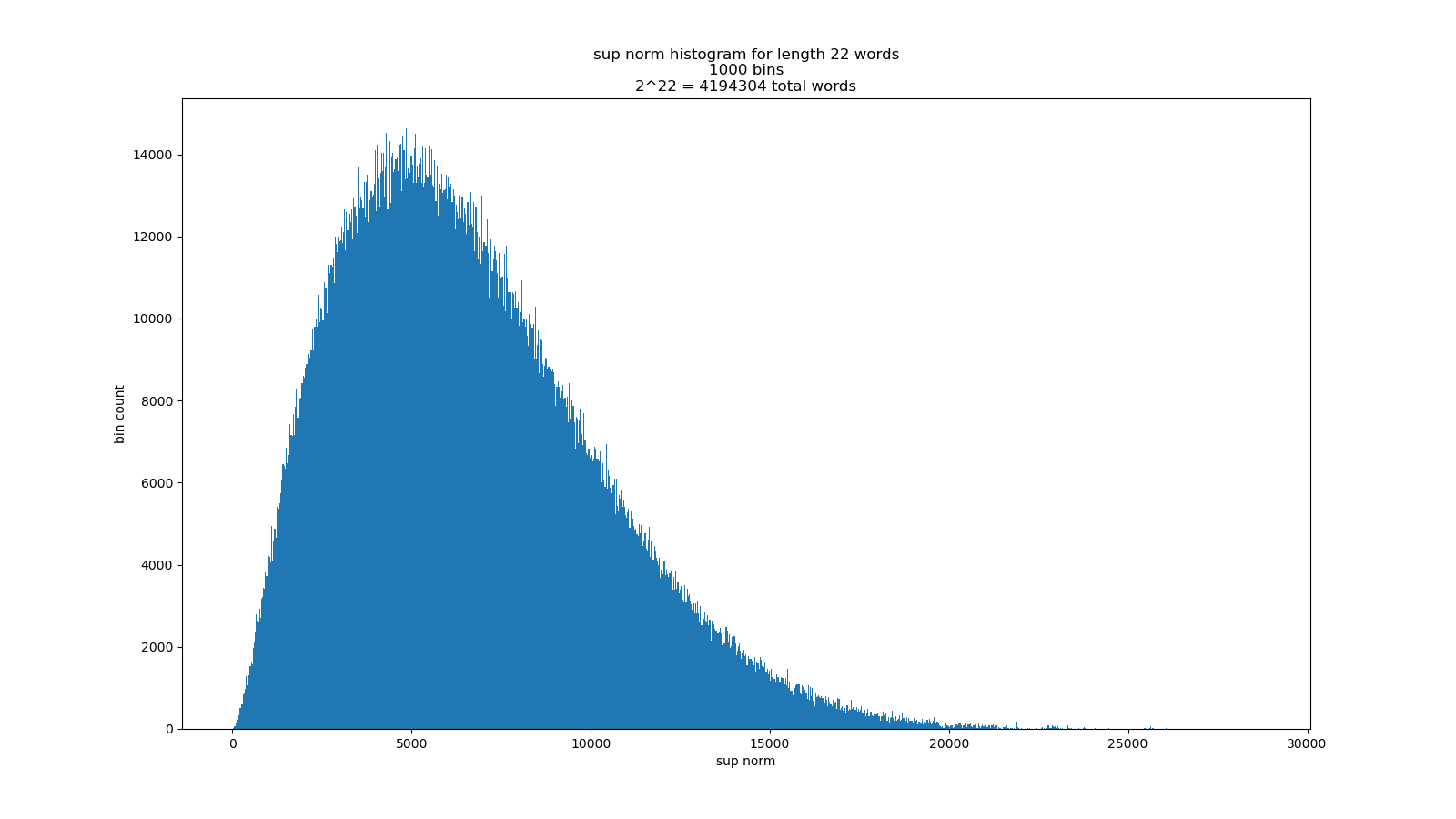}
\includegraphics[width=0.3\textwidth]{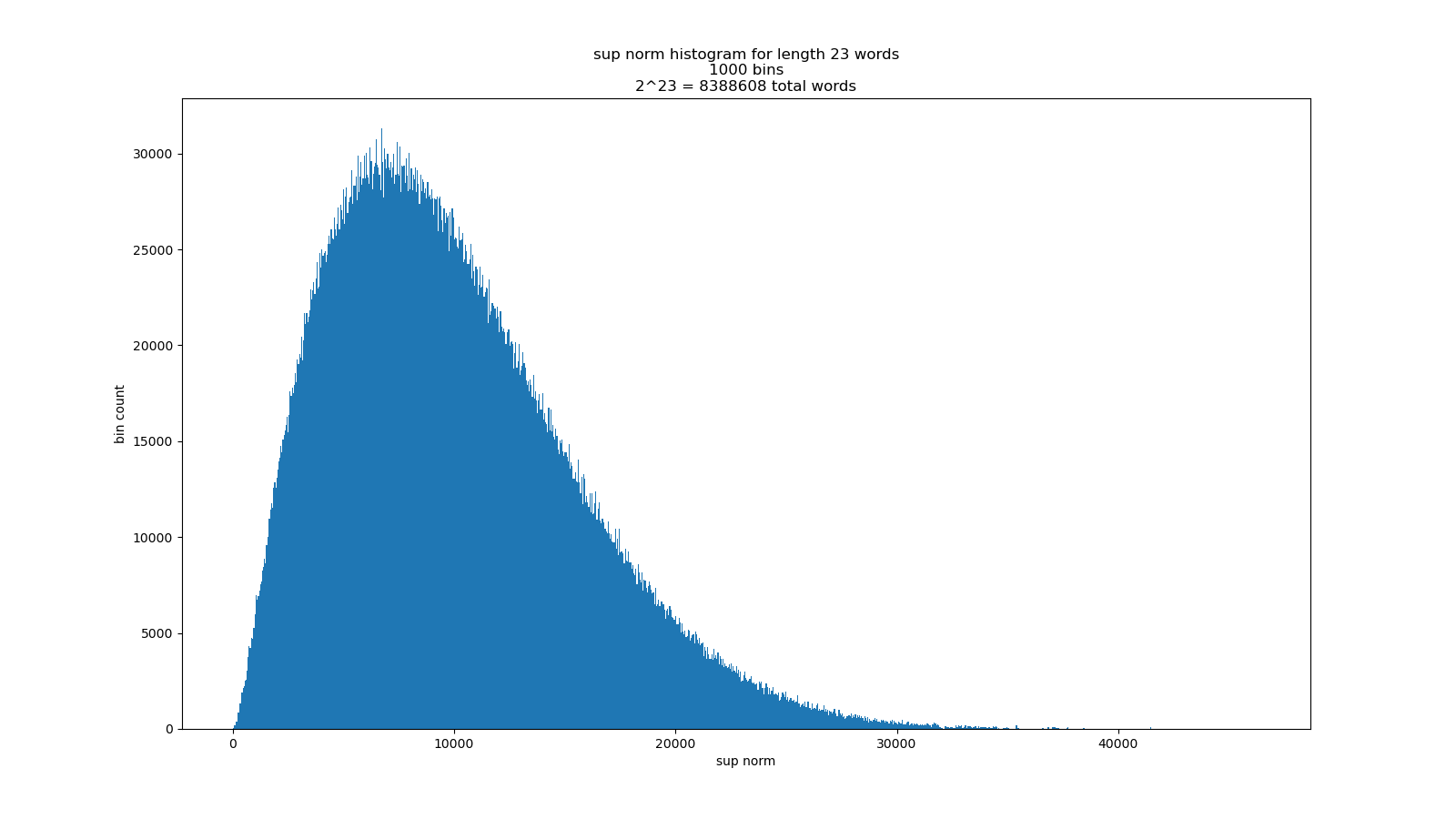}
\includegraphics[width=0.3\textwidth]{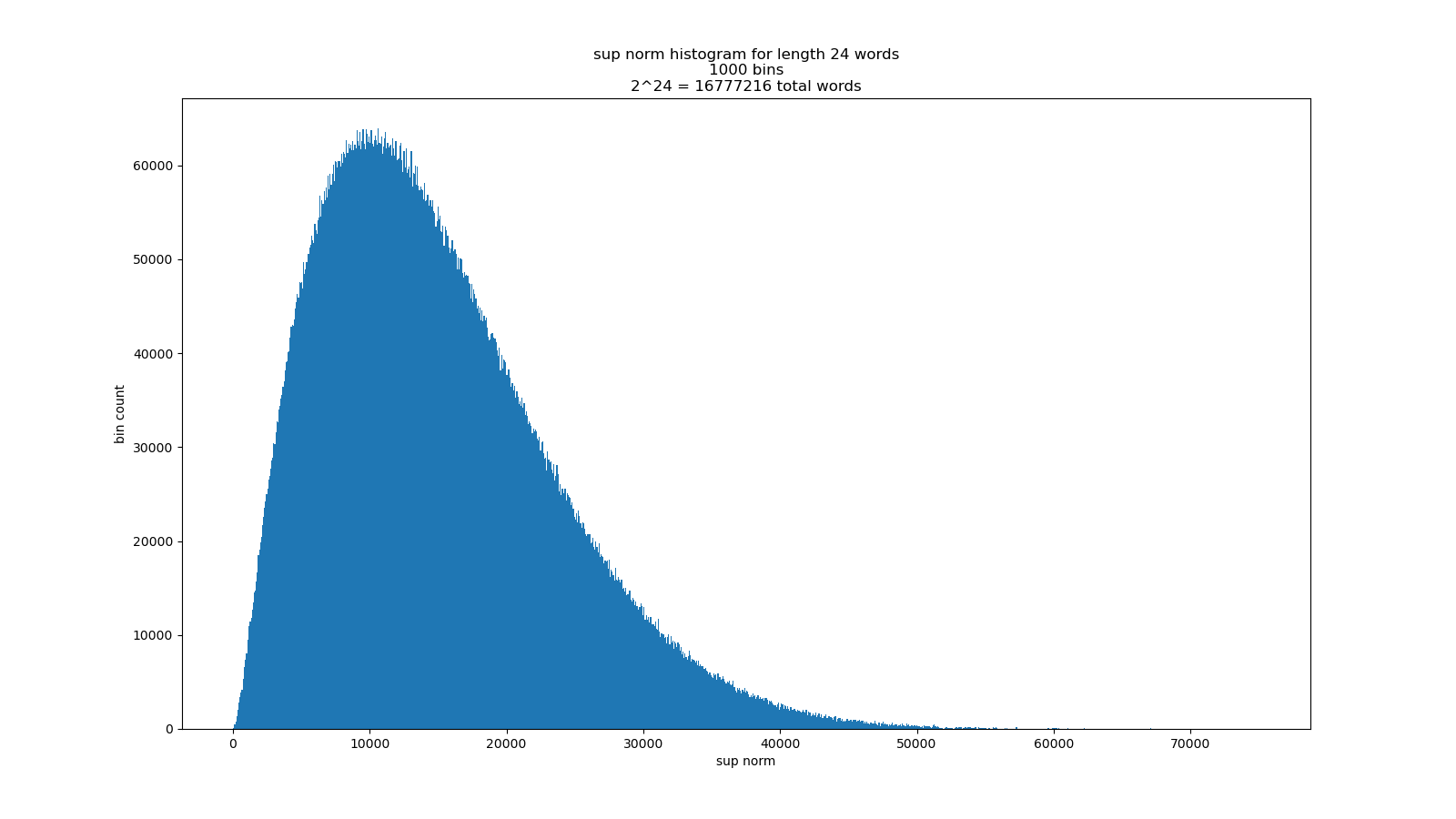}
\caption{$L^{\infty}$ distributions for words of length 22, 23, and 24.  Histograms with 1000 bins of equal size.}
\label{fig:supnormdist}
\end{center}
\end{figure}

Of course, for suitable parameters to be chosen, the joint distribution
of the traces and of \(\|M\|_{\infty}\) needs more detailed analysis.
For a small experiment, \autoref{fig:supnormdist} shows the
\(L^{\infty}\) distribution for words of length \(15\), \(20\), and
\(25\).

\begin{figure}[h]
\begin{center}
\includegraphics[width=0.3\textwidth]{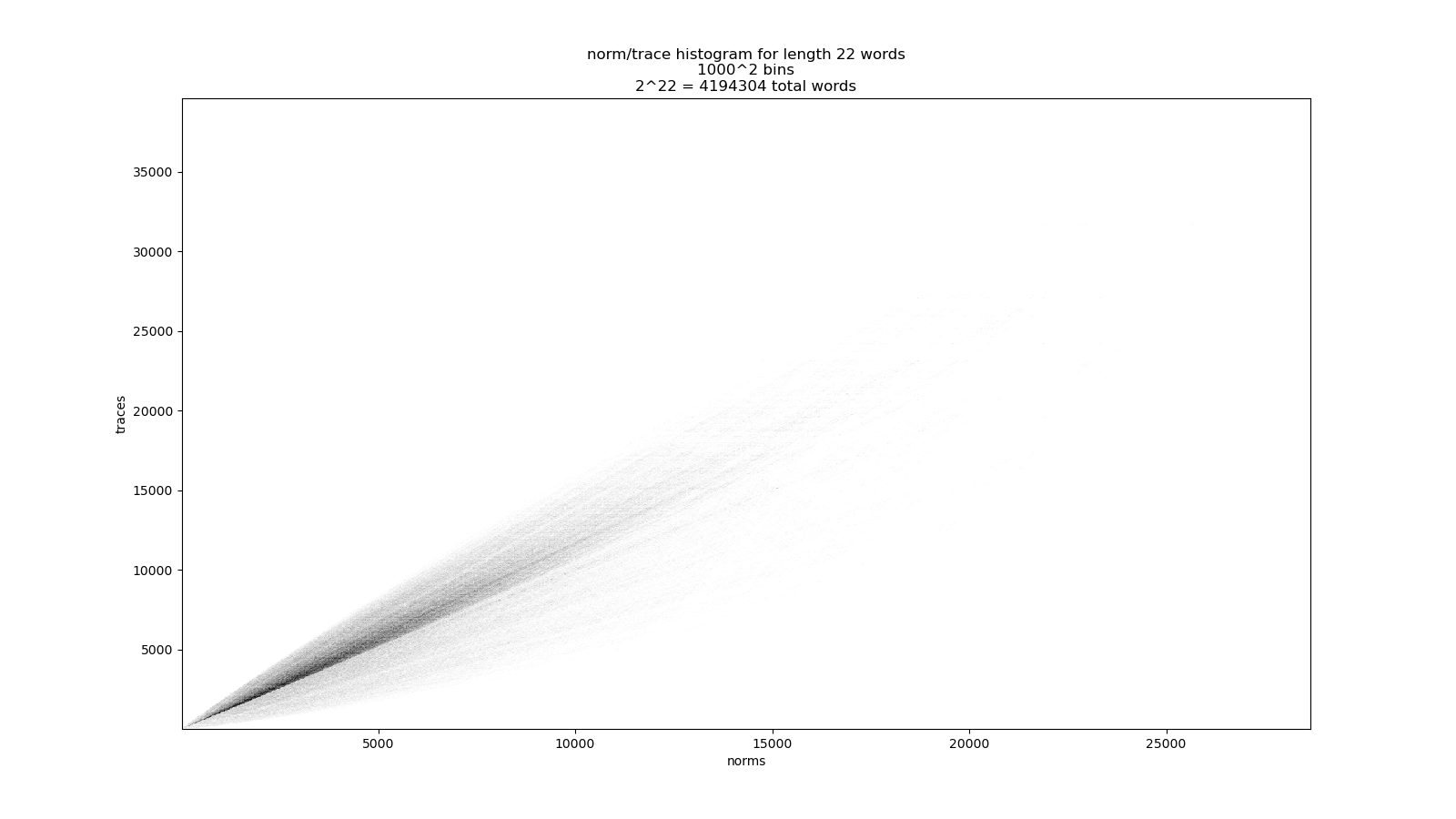}
\includegraphics[width=0.3\textwidth]{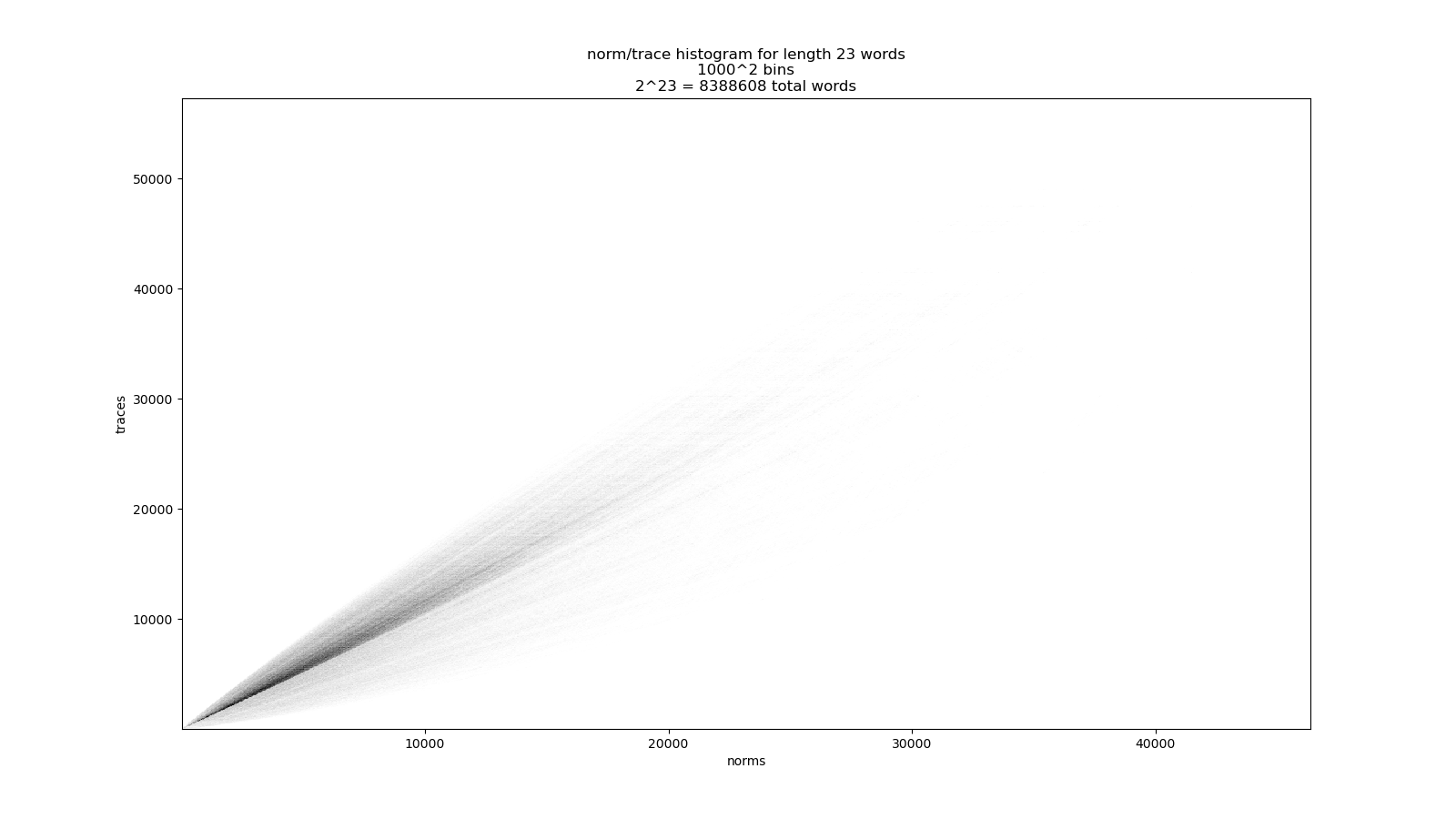}
\includegraphics[width=0.3\textwidth]{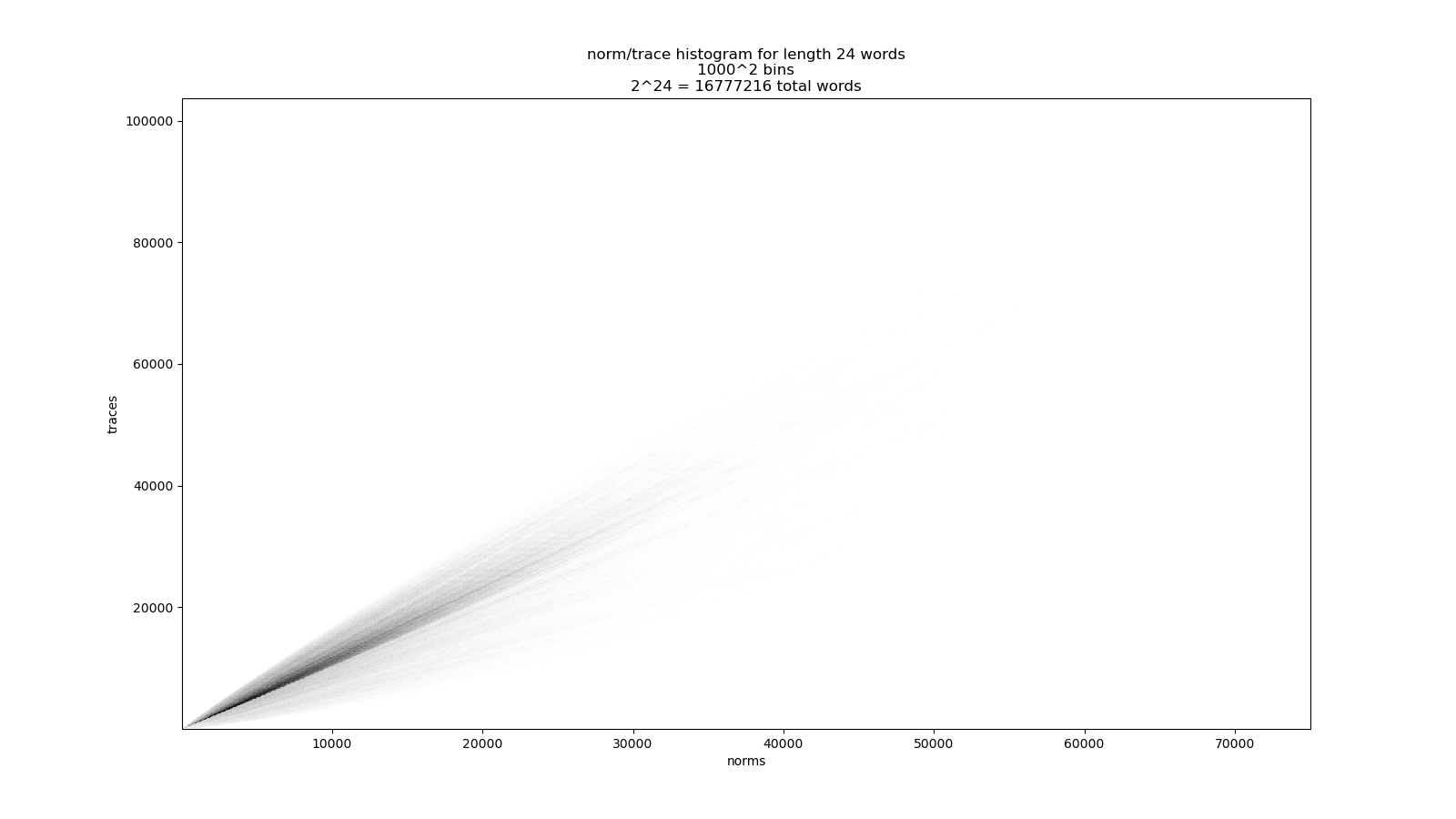}
\caption{Joint distributions for words of length 22, 23, and 24.  Histgrams with 1000 bins in each dimension.}
\label{fig:joint}
\end{center}
\end{figure}

\autoref{fig:joint} gives 2D histograms of the joint distribution for
lengths 22, 23, and 24. The obvious linear relationship seems to be
\(\mathsf{tr}(M)\approx \frac{9}{8}\|M\|_{\infty}\).

To state the question more plainly for reference:

\begin{quote}
What is the distribution of
\(\{(\mathsf{tr}(M), \|M\|_{\infty}) : M\in SL_2^{(k)}(\mathbb{N})\}\)?
\end{quote}

To repeat ourselves, a modest increase in \(k\) to uniformize the trace
distribution and decrease the modular reduction bound \(m\) should
increase both the security and practicality of the scheme.

\hypertarget{questions}{%
\section{Questions}\label{questions}}

We summarize some of the mathematical questions raised above, both for
security and practicality of the scheme.

\begin{enumerate}
\def\labelenumi{\arabic{enumi}.}
\tightlist
\item
  What is the distribution of
  \(\{\mathsf{tr}(M) : M\in SL_2^{(k)}(\mathbb{N})\}\)?
\item
  What is the distribution of
  \(\{\|M\|_{\infty} : M\in SL_2^{(k)}(\mathbb{N})\}\)?
\item
  What is the joint distribution of the two quantities above?
\item
  What is the difficulty (or even a more precise algebraic statement) of
  the secret key search problem, i.e.~finding a uniformly random
  \(S\in GL_2(\mathbb{Z}/m\mathbb{Z})\) given a known
  \(M\in SL_2^{(k)}(\mathbb{N})\) and a known element
  \(S^{-1}\widetilde{M}S\) of the conjugacy class of its ``restriction
  of scalars'' (trivially increasing the dimensions from \(2\times 2\)
  to \(2n\times 2n\))?
\end{enumerate}

Regarding the first three questions, there is the obvious binomial
distribution on the ``abelianized'' structure, i.e.~in terms of the
number of instances of \(L\) and \(R\) in a given word, but the trace
and \(L^{\infty}\) distributions depend on the non-commutativity,
e.g.~more ``switching'' activity increases the trace and maximum matrix
entry. However, in lieu of more detailed information on the actual
distributions, there may be useful bounds on the trace and sup norm as a
function of the relative frequency of \(L\) or \(R\) in a given word.
Such bounds could also greatly reduce the computation needed for
sampling of \(G_0\), \(G_1\), \(\mu\) (i.e.sample based on proportions
of \(L\) and \(R\) before matrix computations).

Towards investigating these questions, we note that the reference
\protect\hyperlink{ref-peter}{{[}3{]}} shows that \[
\lim_{x\to\infty}\frac{1}{x}\left|\{n\leq x: \phi(n)\in(a,b]\}\right|=\int_a^b\delta(t)dt \quad (\text{for }n\geq 3)
\] for some smooth probability distribution \(\delta\), where \[
\phi(n)=\frac{\Phi(n)}{n\log n}, \quad \Phi(n) = |\{M\in SL_2(\mathbb{N}) \ : \ \mathsf{tr}(M)=n\}|,
\] and \protect\hyperlink{ref-kallies}{{[}4{]}} gives an asymptotic for
the summatory function \[
\Psi(N)=\sum_{3\leq n\leq N}\Phi(n)\sim \frac{6}{\pi^2}N^2\log N
\] with further refinements in \protect\hyperlink{ref-boca}{{[}5{]}},
\protect\hyperlink{ref-ustinov}{{[}6{]}}.

\hypertarget{final-thoughts-on-security}{%
\section{Final thoughts on security}\label{final-thoughts-on-security}}

The scheme outlined above doesn't satisfy basic indistinguishability for
two obvious reasons. One is the lack of randomness, which can easily be
remedied, say by including a random mask \(\rho\) with the ciphertext
and encrypting masked plaintext \(\mu\oplus\rho\). The second is the
trace distinguisher, with potential mitigations discussed above
(restricting the range of allowed traces). For the purposes of a KEM, a
good lower bound on the entropy loss due to knowledge of the trace
should suffice.

One nice aspect of the decryption is that it can detect malformed
ciphertext. If presented with ciphertext that is too short, too long, or
built from the wrong generators, etc., the Euclidean algorithm will not
reduce the ciphertext to the identity in the correct number of steps,
and the decryption algorithm can respond with explicit or implicit
rejection.

We also note that the ciphertext is malleable to some extent. Given
valid ciphertext, one can attempt to remove and replace bits from the
beginning and end of the ciphertext (probability \(2^{-k}\) of guessing
\(k\) bits to remove). This could be mitigated by random padding,
i.e.~only the ``middle'' of the plaintext is considered relevant.

We welcome any and all to attack and improve the scheme or further
develop our admittedly weak security analysis.

\hypertarget{references}{%
\section*{References}\label{references}}
\addcontentsline{toc}{section}{References}

\hypertarget{refs}{}
\begin{CSLReferences}{0}{0}
\leavevmode\vadjust pre{\hypertarget{ref-matrix}{}}%
\CSLLeftMargin{{[}1{]} }%
\CSLRightInline{R. Geraud-Stewart and D. Naccache, {``New public-key
cryptosystem blueprints using matrix products in \(\mathbb F_p\).''}
Cryptology ePrint Archive, Paper 2023/1745, 2023.Available:
\url{https://eprint.iacr.org/2023/1745}}

\leavevmode\vadjust pre{\hypertarget{ref-blockinv}{}}%
\CSLLeftMargin{{[}2{]} }%
\CSLRightInline{T.-T. Lu and S.-H. Shiou, {``Inverses of 2 × 2 block
matrices,''} \emph{Computers \& Mathematics with Applications}, vol. 43,
no. 1, pp. 119--129, 2002, doi:
\url{https://doi.org/10.1016/S0898-1221(01)00278-4}.}

\leavevmode\vadjust pre{\hypertarget{ref-peter}{}}%
\CSLLeftMargin{{[}3{]} }%
\CSLRightInline{M. Peter, {``The limit distribution of a number
theoretic function arising from a problem in statistical mechanics,''}
\emph{Journal of Number Theory}, vol. 90, no. 2, pp. 265--280, 2001,
doi: \url{https://doi.org/10.1006/jnth.2001.2666}.}

\leavevmode\vadjust pre{\hypertarget{ref-kallies}{}}%
\CSLLeftMargin{{[}4{]} }%
\CSLRightInline{J. Kallies, A. Özlük, M. Peter, and C. Snyder, {``On
asymptotic properties of a number theoretic function arising out of a
spin chain model in statistical mechanics,''} \emph{Communications in
Mathematical Physics}, vol. 222, no. 1, pp. 9--43, Aug. 2001, doi:
\href{https://doi.org/10.1007/s002200100495}{10.1007/s002200100495}.}

\leavevmode\vadjust pre{\hypertarget{ref-boca}{}}%
\CSLLeftMargin{{[}5{]} }%
\CSLRightInline{F. P. Boca, {``Products of matrices and and the
distribution of reduced quadratic irrationals,''} \emph{Journal für die
reine und angewandte Mathematik}, vol. 2007, no. 606, pp. 149--165,
2007, doi:
\href{https://doi.org/doi:10.1515/CRELLE.2007.038}{doi:10.1515/CRELLE.2007.038}.}

\leavevmode\vadjust pre{\hypertarget{ref-ustinov}{}}%
\CSLLeftMargin{{[}6{]} }%
\CSLRightInline{A. V. Ustinov, {``Spin chains and arnold's problem on
the gauss-kuz'min statistics for quadratic irrationals,''}
\emph{Sbornik: Mathematics}, vol. 204, no. 5, p. 762, Jan. 2013, doi:
\href{https://doi.org/10.1070/SM2013v204n05ABEH004319}{10.1070/SM2013v204n05ABEH004319}.}

\end{CSLReferences}

\end{document}